# Computational Irreducibility and Computational Analogy

## HERVE ZWIRN

*UFR de Physique (LIED, Université Paris 7), CMLA (ENS Cachan, France) & IHPST(CNRS, France)*

*herve.zwirn@gmail.com*

October 2013

## ABSTRACT

In a previous paper [21], we provided a formal definition for the concept of computational irreducibility (CIR), i.e. the fact for a function $f$ from $\mathbf{N}$ to $\mathbf{N}$ that it is impossible to compute $f(n)$ without following approximately the same path than computing successively all the values $f(i)$ from $i$=1 to $n$. Our definition is based on the concept of E-Turing machines (for Enumerating Turing Machines) and on the concept of approximation of E-Turing machines for which we also gave a formal definition. We precise here these definitions through some modifications intended to improve the robustness of the concept. We introduce then a new concept: the Computational Analogy and prove some properties of computationally analog functions. Computational Analogy is an equivalence relation which allows partitioning the set of computable functions in classes whose members have the same properties regarding to their computational irreducibility and their computational complexity.

## 1. Introduction

The notion of Computational Irreducibility (CIR) seems to have been first put forward by Wolfram. Given a physical system whose behavior can be calculated by simulating explicitly each step of its evolution, is it always possible to predict the outcome without tracing each step? Is there always a shortcut to go directly to the $n^{\text{th}}$ step? Wolfram conjectured [16, 17, 18] that in most cases the answer is no. While many computations admit shortcuts that allow them to be performed more rapidly, others cannot be sped up. Computations that cannot be sped up by means of any shortcut are called computationally irreducible.

This question has been widely analyzed in the context of cellular automata by Wolfram [15, 17]. A cellular automaton is computationally irreducible if in order to know the state of the system after $n$ steps there is no other way than to evolve the system $n$ times according to the equations of motion. The intuition behind this definition is that there is no other way to reach the $n^{\text{th}}$ state than to go through the $(n$-1) previous ones.

In this context, Israeli and Goldenfeld in [9] have shown that some automata that are apparently computationally irreducible have nevertheless properties that are predictable. But these properties are obtained by coarse graining and don't account for small scales details.

Moreover some automata (rule 30 for example) seem to be impossible to coarse grain.

Reisinger et al. in [14] show that computational irreducibility seems to be contingent upon the representation of a given problem. To do so, they consider a game for which the initial rules are computationally reducible and they build an isomorphic representation leading to a process that appears to be computationally irreducible. As they notice, a more definitive claim would be to take one of Wolfram's computationally irreducible cellular automata, formulate an isomorphic representation of it, and then determine whether transition rules of the equivalent system are computationally reducible.

Whatever the answers to the questions raised by Israeli and Goldenfeld or by Reisinger et al. are, what is of interest for us in this paper is to provide a robust formal definition of the very concept of computational irreducibility which is lacking. Indeed, as we explained in [21], Wolfram's intuition needs to be rigorously formalized since stated as above, it is not robust. There are two underlying intuitions that seem to be equally important in the concept of CIR. The first one is the question of the speed of computation. If a process is CIR then it should not be possible to compute its $n^{\text{th}}$ state in a time shorter than the time needed to compute successively the $(n$-1) previous states before computing the $n^{\text{th}}$. The second one, is even more demanding. After all, it could well be possible that the time to compute the $n^{th}$ state be not shorter than the sum of the times needed to compute successively all the previous states but that the computation of the $n^{th}$ state doesn't need to really go through the computation of theses states. But for a process to be CIR, the necessity to actually compute these previous states is required. Of course, the second condition implies the first one. In the following, we will address both conditions.

In [21], we provided a first formal definition for the concept of computational irreducibility which we re-expressed in the more general framework of functions $f$ from **N** to **N** as the fact that it is impossible to compute $f(n)$ without following approximately the same path than computing successively all the values $f(i)$ from $i=1$ to $n$. Our definition is based on the concept of E-Turing machines (for Enumerating Turing Machines) and on the concept of approximation of E-Turing machines for which we also gave a formal definition.

In the present paper, we precise these definitions and bring some modifications intended to improve the robustness of the concept. We refer the reader to the original paper for the motivations of the initial definitions. Here, we also introduce a new concept: the Computational Analogy.

In the part 1, we justify the computation model we use throughout this paper. In the part 2, we precise the definition of the E-Turing machines and their approximations and we give more details on the definition of the concept of Computational Irreducibility. In the part 3, we introduce Computational Analogy, discuss its meaning and prove some theorems for functions that are computationally analog, relatively to their computational irreducibility and their computational complexity.

## 2. Part 1 : The computational model

In this paper, we adopt the computational model of Turing machines [6, 7, 8, 13] with $k \geq 2$ tapes. So, let's begin by justifying our choice to use the $k$-tape Turing machines as a good computational model. We are looking for a general model of computation allowing to deal with the questions of efficiency and of speed of computation in a robust way. It is well known that the model of Turing machines is a powerful though very fundamental model of computation. The main point with the Turing machines model is that it is very simple and that through the Church-Turing thesis, it allows the computation of any computable function. Several kinds of Turing machines exist depending on the number of tapes they have. While they are all equivalent regarding the functions they allow to compute, they are not equivalent regarding the speed of computation. For example, the problem of deciding if a string is a palindrome is $O(n^2)$[1] in the 1-tape Turing machines model and $O(n)$ in the 2-tape Turing machines model [7, 13]. Is increasing the number of tapes allowing to improve without limit the speed of the computation of a given problem? This answer is no. A first result [13] says that we can't expect more than a quadratic saving through allowing an arbitrary number of tapes.

**Theorem 2.1.** *Given any $k$-tape Turing machine* M *operating within time* $T(n)$*, it is possible to construct a 1-tape Turing machine* M' *operating within time* $O(T(n)^2)$ *and such that for any input x,* $M(x)=M'(x)$*.*

The meaning of this result is that the best $k$-tape machine that can be designed for doing a computation will never operate in less that $\sqrt{T(n)}$ if the best 1-tape Turing machine doing the same computation operates in a time $T(n)$.

A second result [13] is known as linear speed-up:

**Theorem 2.2.** *For any $k$-tape Turing machine* M *operating in time* $T(n)$ *there exists a $k'$-tape Turing machine* M' *operating in time* $f(n)=\varepsilon T(n)+n$ *(where $\varepsilon$ is an arbitrary small positive constant) which simulates* M.

This linear speed-up means that the main aspect of complexity is captured through the function $T(n)$ irrespectively of any multiplicative constant. **DTIME**$(T(n))$ is the class of functions[2] computable by a $k$-tape Turing machine in $T(n)$ steps. This result means that **DTIME**$(T(n)) = $ **DTIME**$(\varepsilon T(n))$ and so it's legitimate to define **DTIME**$(T(n))$ as the class of functions computable by a Turing machine in $O(T(n))$ steps. If a function $f$ is computable in time $T(n)$ and $\log(f(n))$ (hence the length of its binary representation) is $o(T(n))$ then $f$ is also computable in time $\varepsilon T(n)$ for every $\varepsilon > 0$.

Hence, in the $k$-tape Turing machines model, the speed of computation can be expressed through the $O(T(n))$ notation which is justified. That is what we will do all over the paper as is usual in the field of computational complexity.

More results about the so called "speed-up theorems" are given in our previous paper [21].

Usually, in the theory of computation, one is only interested in knowing if a function is computable and if so, in knowing the computational complexity of getting the output from the input. What is done during the computation is rarely considered and, except for the person writing the program itself, the Turing machine is a kind of black box furnishing an output from an input. But in this paper, we are interested in a particular aspect of computation that is not often addressed: the intermediate results. As we stated in the introduction, a cellular automaton is computationally irreducible if in order to know the state of the system after $n$ steps there is no other way than to evolve the system $n$ times according to the equations of motion. Similarly for a function, to be CIR means that the computation of $f(n)$ requires the previous computation of all the $f(i)$ for $i < n$. CIR functions are defined not by an explicit formula giving directly the value of $f(n)$ from the value of $n$ but by recursive rules giving the way to go from $f(i)$ to $f(i+1)$[3]. So, following these rules, the computation of $f(n)$ starts by the computation of $f(1)$ followed by the computation of $f(2)$ from $f(1)$ then of $f(3)$ from $f(2)$ and so on, till the computation of $f(n)$ from $f(n-1)$. So, in order to be able to characterize that sort of computation, our computational model should allow identifying the intermediate computation steps. For that, we will consider special 3-symbols (0, 1, #) Turing machines such that each of these intermediate results will be successively written on the output tape with the symbol "#" written at its left. More precisely, a program that follows a recursive rule for computing step by step through the iteration of the same rule "knows" when it switches to the next iteration. What we demand in our specific model of computation is that the intermediate result which is the input of the next iteration be written on the output tape at the right of the symbol "#". The final result will appear on the output tape at the right of the last symbol "#". The output tape will be a one way tape (i.e. the head will be allowed to go only in the right direction). We'll see throughout the paper why this kind of special Turing machines is useful for our purpose[4].

---

[1] We refer the reader to the appendix A at the end of the paper for the definition of the standard asymptotic notations.
[2] More precisely the class of decision problems.

In the following $f$, $g$, $h$, F, G, H will always be functions from **N** to **N** and M, P, Q will always be Turing machines as described above.

## 3. Part 2 : The Computational Irreducibility

Given a Turing machine M computing $f(n)$ in time $T(M(n))$, let's denote by $R_{n,1}$, …, $R_{n,i}$, …, $R_{n,T(M(n))}$ the content of the output tape of M during the computation of $f(n)$ after 1 step of computation, …, $i$ steps of computation and $T(M(n))$ steps of computation. So $(R_{n,1}, …, R_{n,i}, …, R_{n,T(M(n))})$ is the sequence of the configurations of the output tape during the computation of $f(n)$.

**Definition 3.1** (E-Turing machine): *A Turing machine* $M_f$ *will be called a* E-*Turing machine for f if:*

*(i)* $M_f$ *computes f (i.e. for every input n,* $M_f$ *computes f(n) and halts). It's important to notice that it is the same Turing machine which on input n computes f(n): f is uniformly computed by* $M_f$.

*(ii) during the computation of f(n), there exist increasing* $k_n(i)$ *for i=1 to n-1, such that f(i) is written on the output tape* $R_{n,k_n(i)}$ *at the right of the last symbol "#".*

A E-Turing machine for a function $f$ (in the following we will always denote $M_f$ such a Turing machine) is a program which, in a certain sense, enumerates the successive values $f(i)$ for $i \le n$. So, during the computation of $f(n)$, $f(1)$ then $f(2)$ and so on until $f(n)$ successively appear on the output tape of $M_f$. It is of course possible to build E-Turing machines for any computable function.

Let $f$ be a computable function. Here are two examples of a E-Turing machine for $f$.

a) Assume first that M is a Turing machine which on every input $n$ computes $f(n)$. Let's now consider the Turing machine $M_f$ which on every input $n$, calls M with input 1 then, when M has computed $f(1)$, write "#" and $f(1)$ on the output tape, calls again M with input 2 and so on until the last call to M with input $n$ and which halts when M has computed $f(n)$ after having written "#" and $f(n)$ on the output tape. $M_f$ is clearly a E-Turing machine for $f$. When computing $f(n)$, $M_f$ will follow exactly the same initial segments than the initial segments followed for all $k < n$ when computing $f(k)$. The computation of $f(n)$ is the continuation of the computation of $f(k)$ for $k < n$. One can also notice that the computation of $f$ for each value $n$ starts from scratch (i.e. the values of $f(k)$ for $k < n$ are not used for computing $f(n)$). This way to build a E-Turing machine is possible for any computable function.

b) Assume now that $f$ is such that it is possible to compute $f(n)$ from $f(n-1)$. Let M' be a Turing machine which on input $f(n-1)$ computes $f(n)$. Let's now consider the Turing machine $M'_f$ which on every input $n$, starts by computing $f(1)$, write "#" and $f(1)$ on the output tape, then calls M' to compute $f(2)$ from the input $f(1)$), write "#" and $f(2)$ on the output tape and so on till $f(n)$. $M'_f$ is a E-Turing machine for $f$. The computation of $f(n)$ by $M_f$ can be seen as the successive computations of $f(i)$ from $f(i-1)$ till reaching $f(n)$. As in the first example, when computing $f(n)$, $M'_f$ follows exactly the same initial segments than the initial segments followed for all $k < n$ when computing $f(k)$. Here again, the computation of $f(n)$ is the continuation of the computation of $f(k)$ for $k < n$.

---

[3] Of course, that doesn't mean that each function defined like that is CIR.
[4] The goal is to be able to distinguish the different results when reading the output tape. Instead of using a special symbol to separate the results, an equivalent method would be to use a self delimiting way to write them.

Because the initial path is the same when computing $f(n)$ and $f(m)$ for $n>m$, these two examples of E-Turing machines can be thought as doing a computation such that on any input $n$, they halt after having run through an initial segment of length $T(M_f(n))$ of one unique infinite virtual computation of $f(i)$ for $i = 1$ to $\infty$. That means also that the $k_n(i)$ are independent of $n$. But this is not necessarily the case for all E-Turing machines.

The computation of $f(n)$ from $f(n-1)$ can be faster than the computation of $f(n)$ from $n$. In this case, $M'_f$ will be much faster than $M_f$. We'll see that this is the case if $f$ is CIR because a Turing machine computing a CIR function $f$ does need to know $f(n-1)$ (or a value that is near in a sense that we will precise) to compute $f(n)$. We give here, some examples of functions more and more "difficult":

- For computing $f(n) = 3^n$ from the input $n$, a Turing machine will go through some of the intermediate values $f(i)$ for $i < n$ but not necessarily all. For instance, $3^{2n}$ can be computed as $3^n$ x $3^n$ and $3^{2n+3}$ will need the computation of $3^{n+1}$ or the computation of $3^n$ and $3^3$. But if $f(n-1)$ is given as input, the computation of $f(n)$ is immediate and fast.

- For computing $f(n) = n!$, a Turing machine will go through $n$ intermediate values if its starts either with $n$ or with $(n-1)!$ as input. Indeed even from $(n-1)!$ it is needed to know $n$ for computing $n!$ and a natural way (but not the only one) to "extract" the value $n$ from $(n-1)!$ is to compute all the increasing values of the factorial function and to count how many have been computed till reaching $(n-1)!$. The computation from $n$ can be done in any possible order since the multiplication of the $n$ first natural numbers can be done from any combination of these numbers. That means that even if a Turing machine computing $n!$ from $n$ will have to perform $n$ operations, it will not necessarily computes all the $k!$ for $k<n$ before. So, it seems that every natural Turing machine computing $n!$ with either $n$ or $(n-1)!$ alone as input will have to perform $n$ operations without having to be necessarily a E-Turing machine. But that will not be the case with the input $(n, (n-1)!)$ from which the computation will be very fast.

- For computing $f(n)$ defined by: "the first bit of the sum of the $k^{th}$ bit of $3^k$ for all $k \le n$", from the input $n$, a Turing machine will go through all the intermediate values $f(i)$ for $i < n$ but will be simply unable to compute $f(n)$ from $f(n-1)$ alone because there is no way to extract the value of $n$ from $f(n-1)$ and this value is needed to compute $f(n)$. So, it seems that every Turing machine computing $f(n)$ with $n$ as input will be a E-Turing machine and $f(n)$ could well be CIR. From the input $(n, f(n-1))$ the computation is fast.

The time $T(M_f(n))$ to compute $f(n)$ with a E-Turing machine $M_f$ is the sum of the times between the apparition on the output tape of $f(i)$ and $f(i+1)$ (from $i=1$ to $n-1$) plus the initial time to get $f(1)$ appearing.

Let's denote $t_i = T(f(i-1) \underset{M_f}{\to} f(i))$ the time between the apparition of $f(i-1)$ and the apparition of $f(i)$ during the computation of $f(n)$ for any $i>i$.

We have $T(M_f(n)) = \sum_{i=1}^{n} t_i$ (we suppose by convention that $t_1$ is the time for $f(1)$ to appear on the output tape). Since $M_f$ is a Turing machine, $t_i$ is the number of steps done by the machine and so is a strictly positive integer. So $T(M_f(n)) \ge n$. But in the following we will be interested only in functions $f$ such that $T(M_f(n)) = \Omega(n \log n)$.

This seems a reasonable assumption and it's obviously true of any function $f$ such that $f(n) \ge n$ since writing an output $n$ in binary or in any other basis $\ge 2$, needs at least a time $\log n$ and a E-Turing machine performs $n$ such operations before halting. So the time for a E-Turing machine computing such a function is necessarily greater than $\Sigma_{i=1 \text{ to } n} \log i = \log(n!) = \Theta(n \log n)$. So $T(M_f(n)) = \Omega(n \log n)$. This is true in particular (see below), for the simulation of a large number of non trivial one dimensional elementary cellular automata with nearest neighbors (which are $\Omega(n^2)$) and in the majority of the simulations of more complex cellular automata (for example Conway's game of life is $\Omega(n^3)$). Of course, we'll consider as well CIR functions for which $f(n) < n$. This is the case of the two candidates given below, at the end of part 2, but it is highly probable that they satisfy nonetheless $T(M_f(n)) = \Omega(n \log n)$.

The question of knowing whether there is an asymptotically optimal program for doing a given computation is a difficult and open question in general. We mean by asymptotically optimal program, a program $p$ such that for any other program $p'$ doing the same computation $T(p(n))=O(T(p'(n)))$. On the one hand, it is well known that the so called Blum's speedup theorem [1] shows that for some decision problems, any program that solves the problem will be much slower than some other program solving the same problem. In these cases, there exists an infinite sequence of programs solving the problem such that each program in the sequence is much faster than the program it follows and (up to a multiplicative constant) there is no asymptotically optimal program. But these problems are artificially constructed to prove the theorem. On the other hand, Levin's Optimal Search Theorem [11] proves that for a wide class of problems there is an asymptotically optimal program. These are problems for which verifying a solution is easy while producing a solution might be difficult. More precisely, these are problems for which the time complexity of checking a solution is asymptotically faster than the time complexity of producing a solution. Now, it is widely thought that no "natural problem" is subject to Blum speedup and that, in general, asymptotically optimal algorithms exist for them. In particular, this is the case for the cellular automata that are the initial source of inspiration for the subject of this paper. Indeed, to show that a program P is asymptotically optimal, it is enough to show that there is a lower bound, say $h(n)$, on the time complexity of any program Q for this problem, $T(Q(n))=\Omega(h(n))$, and to prove that $T(P(n))=O(h(n))$. In this case, P is an asymptotically optimal program. For example, in the case of the simulation of non trivial one dimensional elementary cellular automata with nearest neighbors, it is clear that any algorithm computing the $n$ initial configurations will have in the worst case to perform in $\Omega(n^2)$ and that there are algorithms performing in $O(n^2)$ (see [21] for details on this point). So, these algorithms will be asymptotically optimal. Hence, any Turing machine representing these algorithms will be an asymptotically optimal program for the given cellular automaton. This is what we call in the following of this paper an Efficient E-Turing machine. In the following, we'll make the assumption that there always exist an asymptotically optimal Turing machine that we will note $M^*_f$ and an Efficient E-Turing machine that we will note $M_f^{eff}$ for any function $f$ we consider. Put differently, let's say that we restrict our scope to the subset of the computable functions set made of functions that satisfy this requirement (which is hopefully a very large subset).

We give now the formal definition of an Efficient E-Turing machine for a function, which will be a fundamental building block for what follows.

**Definition 3.2** (Efficient E-Turing machine): *We will say that a* E-*Turing machine* $M_f^{eff}$ *for f is an efficient* E-*Turing machine for f if for any other* E-*Turing machine* $M_f$ *for f:* $T(M_f^{eff}(n)) = O(T(M_f(n)))$ *i.e. there are constants $c > 0$, $n_0 > 0$ such that $\forall n > n_0$, $T(M_f^{eff}(n)) \leq cT(M_f(n))$.*

As explained above, the intuition is that asymptotically it is not possible for a E-Turing machine to compute faster than an efficient E-Turing machine.

It's clear from the definition that for any two efficient E-Turing machines $M_f^{eff}$, $M'^{eff}_f$, and for any two asymptotically optimal Turing machine $M^*_f$, $M'^*_f$, we have: $T(M_f^{eff}(n))=\Theta(T(M'^{eff}_f(n)))$ and $T(M^*_f(n))=\Theta(T(M'^*_f(n)))$. So for any function H, $H(n)=O(T(M_f^{eff}(n)))$ is equivalent to $H(n)=O(T(M'^{eff}_f(n)))$ and $H(n)=O(T(M^*_f(n)))$ is equivalent to $H(n) = O(T(M'^*_f(n)))$. In the following, $M_f^{eff}$ will always denote an efficient E-Turing machine for $f$ and $T(M_f^{eff}(n))$ will denote the time for an efficient E-Turing machine to compute $f(n)$. $M^*_f$ will always denote an asymptotically optimal Turing machine computing $f$ and $T(M^*_f(n))$ will denote the time for an asymptotically optimal Turing machine to compute $f(n)$. According to what is said above, there will be no need to precise which particular efficient E-Turing machine or which asymptotically optimal Turing machine is considered.

We recall that in the following we always suppose that there exist an asymptotically optimal Turing machine $M^*_f$ and an efficient E-Turing machine $M_f^{eff}$ for $f$.

**Definition 3.3** (approximation of a E-Turing machine): *A Turing Machine* M *will be said to be a* P-*approximation[5] of a* E-*Turing machine for f if and only if there are a function* F *such that* $F(n)=O(T(M^*_f(n))/n)$ *and a Turing machine* P *such that for every n:*

*(i) on input n,* M *computes a result $r_n$ such that* P *computes $f(n)$ from n and $r_n$ in a number of steps* $F(n)$ *and halts.*

*(ii) during the computation, there exist non decreasing $k_n(i)$ for i=1 to n-1, such that a result $r'_{n,i}$ is written on the output tape* $R_{n,k_n(i)}$ *at the right of the last symbol "#"and that* P *computes $f(i)$ from n, i and $r'_{n,i}$ in a number of steps* $F(i)$ *and halts[6].*

Actually, if we note $r_n = r'_{n,n}$, P computes always from the triplet $(n, i, r'_{n,i})$ here abbreviated en $n, r_n$ when $i=n$.

Intuitively, an approximation of a E-Turing machine for $f$ is a Turing machine doing a computation that is near the computation made by a E-Turing machine for $f$.

Let's notice that each E-Turing machine for $f$ is of course an approximation of a E-Turing machine for $f$. The associated Turing machine P is simply the identity (a Turing machine which computes $n$ from the input $n$)[7].

An approximation P of a E-Turing machine for $f$ can be a E-Turing machine for $r$ if the $r'_{n,i}$ don't depend on $n$ and if $r'_i = r_i$ for all $i$ (that means that the intermediate results are the values actually computed by P). But it is not necessarily always the case. In particular, it can happens that the intermediate results $r'_{n,i}$ from which P computes $f(i)$ are different for different values of $n$. In this case, the path that M follows for computing $r_n$ is different for different values of $n$ and the $r_i$ for $i<n$ are not necessarily computed.

The concept of approximation of a E-Turing machine for $f$ is actually a concept obtained from the concept of

---

E-Turing machine by relaxing the constraints of the definition along three dimensions. The first one is the fact that on input $n$ an approximation doesn't compute exactly $f(n)$ but a value $r(n)$ such that it is possible to go from $r(n)$ to $f(n)$ through a very short computation. The second one is that the intermediate results don't need to be all the $f(i)$ for $i<n$ but values $r_{n,i}$ from which it is possible to compute $f(i)$ through a very short computation and the third one is that it is even not necessary that the intermediate values be the same on every computation for different $n$.

Another point to notice is that we don't claim that it is necessary to be able to build the Turing machine P which is associated to an approximation through an effective mean. We only ask that such a machine exists.

We can intuitively justify the value chosen for F($n$). F($n$) is the time that the computation of $f(n)$ takes from the value $r_n$ that is computed by the approximation. We have in mind the case of CIR functions for which computing $f(n)$ demands to compute all the previous values. For these functions, since we want $r_n$ to be "near" in a certain sense of $f(n)$, the time to go from $r_n$ to $f(n)$ must be very short compared to the time to compute $f(n)$ from $n$ and at most comparable to the time to compute $f(n)$ from $f(n-1)$. That's the reason why F($n$)= O(T(M*$_f(n)$)/$n$). Indeed, if $f$ is CIR, we'll see that this is the average time to compute $f(n)$ from $f(n-1)$. The factor $1/n$ in T(M*$_f(n)$)/$n$ takes into account the fact that there are $n$ necessary phases to compute $f(n)$ with a E-Turing machine for $f$ and that we want P to compute in a time shorter or equal to each one of these phases.

Another way to understand the value of F($n$), coming from the picture of cellular automata, is to think that $r_n$ is "near" $f(n)$ (and then the computation of $f(n)$ from $r_n$ is fast) if there are only a bounded number of operations to perform on some bits of $r_n$ to go from $r_n$ to $f(n)$. Indeed, in this framework, a bit of $f(n)$ or of $r_n$ is a cell of the cellular automaton. That means that F($n$) is O($l(r_n)$) where $l(r_n)$ is the length of $r_n$. A reasonable assumption is that the length of $r_n$ should not exceed much the length of $f(n)$ so $l(r_n) = $ O(log$f(n)$). That means that F($n$) is O(log$f(n)$). Now as we saw before T(M$_f^{eff}(n)$) = $\Omega$($n$ log$f(n)$) so log$f(n) = $ O(T(M$_f^{eff}(n)$)/$n$) then F($n$) must be O(T(M$_f^{eff}(n)$)/$n$). Now for CIR functions, we anticipate that T(M$_f^{eff}(n)$) = $\Theta$(T(M*$_f(n)$)) so F($n$) = O(T(M*$_f(n)$)/$n$) is equivalent to F($n$) = O(T(M$_f^{eff}(n)$)/$n$). For functions that are not CIR and, on the opposite, satisfy T(M*$_f(n)$) = o(T(M$_f^{eff}(n)$))), the value F($n$) = O(T(M*$_f(n)$)/$n$) is the smaller of the two.

Is it possible to be more demanding and to ask that F($n$) be smaller than that ? The answer is "no" as it is easy to see on the example of one dimension cellular automata. F($n$) is the time for P to compute $f(n)$ from $r_n$ so, in order for P to read $r_n$ and to write $f(n)$, F($n$) must be at least equal to $l(f(n))$. For non trivial automata, $l(f(n)) = $ O($n$). If these automata are CIR, then T(M$_f^{eff}(n)$) = $\Theta$(T(M*$_f(n)$)) = O($n^2$) and so F($n$) = O(T(M*$_f(n)$)/$n$) = O($n$). One can see that demanding a smaller value for F($n$) would result in the fact that no machine P can exist since the time to write $f(n)$ is $\Omega$($n$). By the way, one can notice that for these automata that are not CIR and for which T(M*$_f(n)$) = o($n^2$), there will be no machine P and hence no approximation of E-Turing machine for these automata. Indeed, in this case F($n$) = O(T(M*$_f(n)$)/$n$) = o($n$) which is too small a value for any P to write $f(n)$. Of course, this is true only for functions such that $l(f(n)) > n$ which is not mandatory. In particular, this is false for trivial automata whose configurations vanish after some iterations or for which the successive configurations are restricted to one cell. So, the above reasoning is not a proof but only an intuitive justification of the value of F($n$).

**Definition 3.4** (Computation of $f(n)$ based on an approximation): *Let* M *be a* P*-approximation of a* E*-Turing machine for f. Let's consider the computation of f(n) done initially through* M *with input n and continued when* M *has computed $r_n$ by* P *which computes f(n) from n and $r_n$ in a time* F*(n) and halts. This computation will be said to be a computation of f(n) based on the* P*-approximation* M.

**Definition 3.5** (Turing machine computing $f$ based on an approximation): *Let* M *be a* P*-approximation of a* E*-Turing machine for f and let's consider the Turing machine* M' *which, for every n, computes f(n) through a computation based on the* P*-approximation* M. M' *will be said to be a Turing machine computing f based on the approximation* M.

If M is a E-Turing machine for $f$, M and M' are identical and M' is of course a E-Turing machine for $f$. Otherwise, M' is also an approximation of a E-Turing machine for $f$. The Turing machine P' associated to M' is the same than P, i.e. P' computes $f(i)$ from $n$, $i$ and $r'_{n,i}$ in a number of steps F($i$), except that for the computation on input $n$, $n$ and $r'_{n,n}$, P' is the identity while P computes $f(n)$.

As shown in theorem 3.1, the important point is that it is possible to build a E-Turing machine for $f$ from any approximation of a E-Turing machine for $f$.

**Theorem 3.1:** *From any* M *approximation of a* E*-Turing machine for f it is possible to build a* E*-Turing machine* M' *for f (we'll call it the daughter of* M) *computing in a time* T(M'($n$)) = $\Theta$(T(M($n$))).

**Proof**: Since M is an approximation of a E-Turing machine for $f$, there are a Turing machine P and a function F associated as mentioned in the definition 3.3. Let's consider the Turing machine built according the following way: on input $n$, M' does exactly the same computation than M but for each $i<n$, after having computed $r_{n,i}$ which computes $f(i)$ through P with input $n$, $i$, $r_{n,i}$ in a time F($i$), writes "#" and $f(i)$ on its output tape and resumes the computation and, at last, computes $f(n)$ from $n$ and $r_n$. It's clear that M' is a E-Turing machine for $f$. M' computes in a time:

$$T(M'(n)) = T(M(n)) + \sum_{i=1}^{n} F(i) + O(1)$$

$$= T(M(n)) + \sum_{i=1}^{n} O(T(M_f^r(i))/i) + O(1)$$

Now it is possible to compute $f(n)$ by M followed by P (that is a computation of $f$ based on the approximation M) so $T(M^*_f(n)) = O(T(M(n)) + F(n))$

$T(M^*_f(n)) = O[T(M(n)) + O(T(M_f^r(n))/n)]$ Hence

$T(M^*_f(n)) = O(T(M(n)))$. Then

$T(M'(n)) = T(M(n)) + \sum_{i=1}^{n} O(T(M(i))/i)]$

Now $\sum_{i=1}^{n} F(i))/i = O(F(n))$ if F is a convex function and $F(n) = \Omega(\log n)^8$. Since any function $O(T(M(i)))$ is a convex function $\Omega(\log n)$, we have

$T(M'(n)) = T(M(n)) + O(T(M(n))) = O(T(M(n)))$

As $T(M(n)) < T(M'(n))$ we get $T(M(n)) = \Theta(T(M'(n)))$

In the following, we will note $\otimes$ this particular form of composition of the two Turing machines M and P that we described above. So M' = P $\otimes$ M. The composition $\otimes$ is defined for a pair (P, M) when the second argument is an approximation of a E-Turing machine for a given function $f$ and the first one is the associated Turing machine computing $f(i)$ from the intermediate results of M. Of course, this composition is not to be confused with the usual composition PoM which runs first the program M and then the program P with the result of the computation of M as input. An important difference is the computation time. The computation time of PoM is the sum of the respective computation times:

$T((PoM) (n)) = T(P(\text{output of } M(n))) + T(M(n))$.

While the computation time of P $\otimes$ M is:

$$T((P \otimes M)(n)) = \sum_{i=1}^{n} T\left(P\left(r_{n,i}\right)\right) + T(M((n)) + O(1)$$

$$= \sum_{i=1}^{n} F(i) + T(M((n)) + O(1)$$

**Theorem 3.2:** *No approximation of a E-Turing machine for $f$ can compute faster than an efficient E-Turing machine for $f$. More precisely, if* M *is an approximation of a E-Turing machine for $f$ then* $T(M_f^{eff}(n)) = O(T(M(n)))$.

**Proof:** Let M' be the daughter of M. Since M' is a E-Turing machine for $f$, $T(M_f^{eff}(n)) = O(T(M'(n)))$. By theorem 3.1 we have $T(M'(n)) = \Theta(T(M(n)))$. So $T(M_f^{eff}(n)) = O(T(M(n)))$.

---

**Theorem 3.3:** *Let* M' *be a Turing machine computing $f$ based on an approximation* M.
*Then:* $T(M'(n)) = \Theta(T(M(n)))$.

**Proof**: M' will compute in a time $T(M'(n))$ such that $T(M(n)) \leq T(M'(n)) \leq T(M(n)) + F(n)$
$= T(M(n)) + O(T(M^*_f(n))/n)$
$= T(M(n)) + O(T(M(n))/n) = O(T(M(n)))$
So $T(M(n)) = \Theta(T(M'(n)))$

In summary, one can say that an approximation of a E-Turing machine for $f$, its daughter and any Turing machine computing $f$ based on this approximation compute all in the same time.

**Definition 3.6** (strongly CIR (resp CIR) function): *A function $f(n)$ from* **N** *to* **N** *will be said to be strongly* CIR (*resp* CIR) *if and only if for any Turing machine* M *computing $f$ there is a* P*-approximation of a* E*-Turing machine for $f$,* M' *such that for every $n$ (resp. for infinitely many $n$), the computation of $f(n)$ by* M *is based on* M'.

The intuition is that if a function is strongly CIR, for each $n$ there is no other way to compute $f(n)$ than to compute before all the values $f(i)$ for $i<n$ (or values that are near in the sense given in the definition of an approximation of a E-Turing machine). There is no shortcut allowing to get directly the value of $f(n)$ without having computed before $f(n-1)$ or a value that is near $f(n-1)$ and so forth for the previous values. If a function is CIR (but not strongly CIR), for infinitely many $n$ there is no other way to compute $f(n)$ than to compute before all the values $f(i)$ for $i<n$ (or values that are near).

The reason why it's useful to introduce this distinction between strongly CIR and CIR can be explained through the following example. Assume that $f$ is strongly CIR. So there is no other way to compute $f(n)$ than to compute before all the values $f(i)$ for $i<n$ (or values that are near) and that is true for every $n$. Let's now consider the function $g$ such that $g(2i-1)=f(i)$ and $g(2i)=1$. It's clear that computing $g$ for any even value is very easy and doesn't imply having to compute any other result before. So, $g$ is not strongly CIR. But, the intuition is nevertheless that $g$ is irreducible in some way. So, the notion of strongly CIR needs to be weakened to cover functions like $g$ and many others that similarly need infinitely often (but not always) to go through the computation of all the previous values for computing them.

**Theorem 3.4:** *If a function $f$ is strongly* CIR *then no Turing machine computing $f$ can compute $f(n)$ faster than an efficient* E*-Turing machine for $f$. So for any Turing machine* M *computing $f$,* $T(M_f^{eff}(n)) = O(T(M(n)))$.

---

**Proof:** If $f$ is strongly CIR, then any Turing machine M computing $f$ is based on an approximation of a E-Turing machine for $f$. Let M' be this approximation. From theorem 3.2, $T(M_f^{eff}(n)) = O(T(M'(n)))$. From theorem 3.3, $T(M(n)) = \Theta(T(M'(n)))$.

So $T(M_f^{eff}(n)) = O(T(M(n)))$.

This result is slightly weakened in the theorem 3.5 if $f$ is simply CIR. In this case, for any Turing machine computing $f$, there are infinitely many values of $f(n)$ that is not possible to compute faster than the computation by an efficient E-Turing machine for $f$.

**Theorem 3.5:** *If a function $f$ is CIR then for any Turing machine M computing $f$ there are constants $c > 0$, $n_0 > 0$ such that $\forall N > n_0$, $\exists n > N$, $T(M_f^{eff}(n)) \leq c\,T(M(n))$.*

**Proof:** If $f$ is CIR, then for any Turing machine M computing $f$ there is a P-approximation of a E-Turing machine for $f$, M', such that for infinitely many $n$, the computation of $f(n)$ by M is based on M'. From theorem 3.2, $T(M_f^{eff}(n)) = O(T(M'(n)))$. So, there are constants $c > 0$, $n_0 > 0$ such that $\forall n > n_0$, $T(M_f^{eff}(n)) \leq c\,T(M'(n))$. But $\forall N$, $\exists n > N$ such that the computation of $f(n)$ by M is based on M'. For such $n$ $T(M(n)) > T(M'(n))$. So for those $n$ that are superior to $n_0$, $T(M_f^{eff}(n)) \leq c\,T(M(n))$.

**Theorem 3.6:** *If a function $f$ is strongly CIR then $T(M^*_f(n)) = \Theta(T(M_f^{eff}(n)))$.*

**Proof:** If $f$ is strongly CIR, then by theorem 3.4 for any Turing machine M computing $f$, $T(M_f^{eff}(n)) = O(T(M(n)))$. So $T(M_f^{eff}(n)) = O(T(M^*_f(n)))$. By definition for any Turing machine M computing $f$, $T(M^*_f(n)) = O(T(M(n)))$. So $T(M^*_f(n)) = O(T(M_f^{eff}(n)))$.

Hence $T(M^*_f(n)) = \Theta(T(M_f^{eff}(n)))$.

Definition 3.7 and theorems 3.4, 3.5 and 3.6 address the two key points of the underlying intuitions for the concept of CIR: the speed of computation and the path followed during the computation.

Let's give two examples of functions that seem to be good candidates to be strongly CIR (but of course, a rigorous proof remains to be found).

1. Let $\mathcal{B} = \{0,1\}$ and $\mathcal{B}^*$ be the set of all finite strings over $\mathcal{B}$, let $\mathcal{L}$ be a recursive language and assume an enumeration of the words of $\mathcal{B}^*$ (for example the index in the length-increasing lexicographic ordering). Define the function $f$ by $f(n)$ is the number of words $w_i$ (for $i \leq n$ in the chosen enumeration) of $\mathcal{B}^*$ in $\mathcal{L}$. Then it seems that, in general, there is no other way to compute $f(n)$ than to decide for each $i \leq n$ if the word $w_i$ belongs or not to $\mathcal{L}$ and to count the number of positive answers.

2. Knowing if an initial configuration of Conway's game of life will be eternal or not is an undecidable problem. So let $f(n)$ be the number of initial configurations of index smaller than $n+1$ in a given enumeration that are still living after $n$ iterations. Here again, it seems that there is no other way to compute $f(n)$ than to test each one of the relevant configurations during $n$ steps and therefore, so doing, to go through the computation of all the $f(i)$ for $i < n$.

## 4. Part 3 : The Computational Analogy

Let M be an approximation of a E-Turing machine for $f$. M computes a function $r$ but is not necessarily a E-Turing machine for $r$. Nevertheless, it is clear that each E-Turing machine for $r$ is an approximation of a E-Turing machine for $f$. But it is possible that no E-Turing machine for $f$ be an approximation of a E-Turing machine for $r$. It would be the case if, while the time to go from $n$, $r(n)$ to $f(n)$ through P is $O(T(M^*_f(n))/n)$, there is no Turing machine able to compute $r(n)$ from $n$, $f(n)$ in a time $O(T(M^*_r(n))/n)$ where $M^*_r$ is an asymptotically optimal Turing machine for $r$. But if one E-Turing machine for $f$ is an approximation of a E-Turing machine for $r$ then every E-Turing machine for $f$ will be an approximation of a E-Turing machine for $r$. In this case, each E-Turing machine for $f$ is an approximation of a E-Turing machine for $r$ and vice versa, each E-Turing machine for $r$ is an approximation of a E-Turing machine for $f$. So it's possible to define a relation of "computational analogy" **CA** (which will be proved to be an equivalence relation):

**Definition 4.1** (Equivalence Relation: Computational Analogy): *$f$ and $g$ will be said to be computationally analog (noted $f$ CA $g$) if:*
*(i) there exists a Turing machine M that is both a E-Turing machine for $f$ and an approximation of a E-Turing machine for $g$*
*(ii) there exists a Turing machine M' that is both a E-Turing machine for $g$ and an approximation of a E-Turing machine for $f$*

That means that there is a Turing machine $P^{f->g}$ which computes $g(n)$ from $n$, $f(n)$ for every $n$ in a time $F(n) = O(T(M^*_g(n)/n))$ (and vice versa). So:

**Theorem 4.1:** *($f$ CA $g$) is equivalent to: there is a Turing machine $P^{f->g}$ which computes $g(n)$ from $n$, $f(n)$ for every $n$ in a time $F(n) = O(T(M^*_g(n)/n))$ and there is a Turing machine $P^{g->f}$ which computes $f(n)$ from $n$, $g(n)$ for every $n$ in a time $G(n) = O(T(M^*_f(n)/n))$.*

In the following, when $f$ CA $g$, we will always denote by $P^{g->f}$ and $P^{f->g}$ these Turing machines.

**Theorem 4.2**: *Let* $M*_f$ *(resp.* $M*_g$*) be an asymptotically optimal Turing machine computing f (resp. g). If f* **CA** *g then* $T(M*_f(n)) = \Theta(T(M*_g(n)))$.

**Proof:** The Turing machine $(P^{f->g} \circ M*_f)$ computes $g$ in a time $T(M*_f(n)) + F(n)$ with $F(n) = O(T(M*_g(n))/n)$. Since $M*_g$ is an asymptotically optimal Turing machines computing $g$, $T(M*_g(n)) = O(T(M*_f(n))) + O(T(M*_g(n))/n)$ so $T(M*_g(n)) = O(T(M*_f(n)))$. The same reasoning with $(P^{g->f} \circ M*_g)$ proves that $T(M*_f(n)) = O(T(M*_g(n)))$. Then $T(M*_f(n)) = \Theta(T(M*_g(n)))$.

**Theorem 4.3**: *Let* $M^{eff}_f$ *(resp.* $M^{eff}_g$*) be an asymptotically optimal Turing machine computing f (resp. g). If f* **CA** *g then* $T(M^{eff}_f(n)) = \Theta(T(M^{eff}_g(n)))$

**Proof:** The Turing machine $P^{f->g} \otimes M^{eff}_f$ which is a E-Turing machine for $g$ computes in a time $T(P^{f->g} \otimes M^{eff}_f) = T(M^{eff}_f(n)) + \sum_{i=1}^{n} O(T(M*_g(i))/i)$ . Since $M^{eff}_g$ is an efficient E-Turing machine $T(M^{eff}_g(n)) = O(T(P^{f->g} \otimes M^{eff}_f(n)))$.

Hence $T(M^{eff}_g(n)) = O(T(M^{eff}_f(n)) + \sum_{i=1}^{n} O(\frac{T(M*_g(i))}{i}))$. Now $\sum_{i=1}^{n} O(T(M*_g(i))/i) = O(T(M*_g(n)))$ (see appendix B).

So $T(M^{eff}_g(n)) = O(T(M^{eff}_f(n))) + O(T(M*_g(n)))$. Now $T(M*_g(n)) = \Theta(T(M*_f(n)))$ by theorem 4.2 and $T(M*_f(n)) = O(T(M^{eff}_f(n)))$ then $T(M^{eff}_g(n)) = O(T(M^{eff}_f(n)))$.

The same reasoning for $f$ shows that $T(M^{eff}_f(n)) = O(T(M^{eff}_g(n)))$.

Hence $T(M^{eff}_f(n)) = \Theta(T(M^{eff}_g(n)))$.

**Theorem 4.4:** *(f* **CA** *g) is equivalent to: any approximation of a* E-*Turing machine for f is an approximation of a* E-*Turing machine for g and vice versa.*

**Proof:** Consider first the direct sense: Let M be a P-approximation of a E-Turing machine for $f$. P computes in a time $O(T(M*_f(n))/n)$. According to theorem 4.1, there is a Turing machine $P^{f->g}$ which computes $g(n)$ from $f(n)$ for every $n$ in a time $F(n) = O(T(M*_g(n))/n)$.

It's then clear that M is a $(P^{f->g} \circ P)$-approximation of a E-Turing machine for $g$ because $(P^{f->g} \circ P)$ computes in a time $O(T(M*_f(n))/n) + O(T(M*_g(n))/n) = O(T(M*_g(n))/n)$ since by theorem 4.2, $T(M*_f(n)) = \Theta(T(M*_g(n)))$. Consider now the reverse sense: a E-Turing machine for $f$ is an approximation of a E-Turing machine for $f$ so it is an approximation of a E-Turing machine for $g$ (and vice versa).

The very meaning of $f$ **CA** $g$ is that $f$ and $g$ share the same approximations of E-Turing machines.

**Theorem 4.5:** **CA** *is an equivalence relation*
**Proof:** This is obvious by theorem 4.4.

The quotient set of the computable functions[9] set by this equivalence relation is made of equivalence classes of computationally analog functions (**CA** functions) that share properties about their computational complexity (their asymptotically optimal programs compute in the same time as well as their efficient E-Turing machines, by theorems 4.2 and 4.3) and their computational irreducibility as we are now going to show.

**Theorem 4.6:** *Assume f* **CA** *g. If f is strongly CIR then g is strongly CIR.*

**Proof:** Let M be a Turing machine computing every $g(n)$. Since $f$ **CA** $g$, there is a Turing machine $P^{g->f}$ which computes $f(n)$ from $n$, $g(n)$ for every $n$ in a time $F(n) = O(T(M*_f(n))/n)$. $(P^{g->f} \circ M)$ is a Turing machine computing every $f(n)$. Now $f$ is strongly CIR, so, there are a Turing machine S which is an approximation of a E-Turing machine for $f$, a Turing machine Q and a function $H(n) = O(T(M*_f(n))/n)$ such that for every $n$, the computation of $f(n)$ made by $(P^{g->f} \circ M)$ is based on S (i.e. is actually the same than the computation of $f(n)$ made by S followed by Q which computes in a time $H(n)$). Since $f$ **CA** $g$, by theorem 4.4, S is also an approximation of a E-Turing machine for $g$. So during the computation of $f(n)$ there are data $r_{n,i}$ (computed by S) appearing successively in an increasing order from $i=1$ to $n$ on the output string of S such that there is a Turing machine Q' that on input $r_{n,i}$, computes $g(i)$ in a number of steps $H'(i)$ (where $H'(n) = O(T(M*_g(n))/n)$). Since $(P^{g->f} \circ M)$ and (QoS) are the same Turing machine, that means that some of these $r_{n,i}$ appear during the computation of M and some appear during the computation of $P^{g->f}$. Let's assume that all the $r_{n,i}$ for $i=1$ to $k$, appear during the computation of M and that all the $r_{n,i}$ for $i=k+1$ to $n$, appear during the computation of $P^{g->f}$ Let's now consider the Turing machine Q'' gotten from Q' through the following change:
- on input $n$, $i$, $r_{n,i}$ for $i=1$ to $k$, Q'' does the same computation than Q' (i.e. computes $g(i)$ in a time $H'(i)$).
- on input $n$, $i$, $r_{n,k}$ for $i=k+1$ to $n$, Q'' starts by computing $r_{n,i}$ then computes $g(i)$ from $r(i)$ as Q' does.

Since $P^{g->f}$ computes $f(n)$ from $n$, $g(n)$ in a time $G(n) = O(T(M*_f(n)/n))$, all the $r_{n,i}$ for $i=k+1$ to $n$, will appear in a time less than $G(n)$. So the computation of $g(i)$ from $n$, $i$, $r_{n,k}$ (for $i=k+1$ to $n$), will be done in a time $H''(i)$ smaller than $G(n) + H'(i)$. Since $G(n) = O(T(M*_g(n))/n)$, which is equal to $O(T(M*_g(n))/n)$ by theorem 4.2, and since $H'(n) = O(T(M*_g(n))/n)$ we get $H''(n) = O(T(M*_g(n))/n)$.

Let's notice that the list of intermediate results $r'_{n,i}$ from which Q'' computes $g(i)$ is the same than the list of $r_{n,i}$ for $i=1$ to $k$ and is equal to $r_{n+k}$ for $i=k$ to $n$. That means that M is based on a Q"-approximation of a E-Turing machine for $g$ (the Turing machine computing all the $r_{n,i}$ for $i=1$ to $k$) and so $g$ is strongly CIR.

**Theorem 4.7:** *Assume f **CA** g. If f is CIR then g is CIR.*
**Proof:** Let M be a Turing machine computing every $g(n)$. Since $f$ **CA** $g$, there is a Turing machine $P^{g->f}$ which computes $f(n)$ from $n$, $g(n)$ for every $n$ in a time $F(n) = O(T(M_f*(n)/n))$. $(P^{g->f}oM)$ is a Turing machine computing every $f(n)$. Now $f$ is CIR so there is an approximation S of a E-Turing machine for $f$ such that for infinitely many $n$ the computation of $f(n)$ by $(P^{g->f}oM)$ is based on S. Let's consider the function $f'$ obtained from $f$ by: $f'(n) = f(p)$ where $p$ is the $n^{\text{th}}$ value for which the computation of $f$ by $(P^{g->f}oM)$ is based on S. It's clear that $f'$ is strongly CIR since for every $n$, the computation of $f'(n)$ is based on the approximation S' which does exactly the same computation than S excepted that on input $n$, S' computes the result that S computes on input $p$ where $p$ is the $n^{\text{th}}$ value for which the computation of $f$ by $(P^{g->f}oM)$ is based on S. Let $g'$ be the function defined similarly from $g$: $g'(n) = g(p)$ where $p$ is the $n^{\text{th}}$ value for which the computation of $f$ by $(P^{g->f}oM)$ is based on S. It's clear that $f'$ **CA** $g'$. So, $g'$ is strongly CIR. Then $g$ is CIR.

## 5. Conclusion

We have provided a formal definition of Computational Irreducibility that clarifies the intuition about this concept and that allows to understand that a function is CIR if there is a class of close paths that it is necessary to follow in order to compute it. In a broad sense, that means that if a function is computationally irreducible, there is only one road (the width of the road being the size of the class of the close paths that one can use) to compute this function. In a way, all these paths have the same length. This explains the fact that it is not possible to go faster than following one these paths to compute the function. We have also defined an equivalence relation between functions that share the same road. Roughly speaking, Computational Analogy allows to get a quotient set which can be viewed as a map of the computable functions set (or at least a large subset of this set whose elements satisfy the conditions for the above concepts to be applicable) for which classes are grouping elements having similar properties relatively to their time of computation and their computational irreducibility.

An open problem is still to prove that one function among the possible candidates is really CIR. The cellular automaton rule 110 which has been shown to be universal (see [4]), the first candidate we mention at the beginning of part 2 or the two other examples we proposed at the end of part 2 are good examples of functions that we would like to prove CIR.

On a more philosophical point of view, Computational Irreducibility can help clarifying the concept of emergence and can be used to understand why certain phenomena appear to be emergent. We have proposed in [19] and [20] that "understanding" a process implies having a mental model of it that we can use to simulate its behavior. Emergent phenomena are effects or properties appearing at the macro level (collective) of a system and that are caused by the micro level (individual) but very difficult and even seemingly impossible to predict even from the complete knowledge of the rules of the micro level. Now if the process running at the micro level is CIR or if the rules leading from the micro level to the macro level are CIR then the global behavior of the system will be neither predictable (without simulating it) nor understandable. In this case, what happens will be seen as "emergent". For example, the fact that some patterns (pulsar, glider, glider gun ...) are usually considered as emergent in Conway's game of life could be explained by the fact that the underlying rules are CIR. Similarly, phenomena that are sometimes interpreted as downward causation could be merely CIR processes interpreted as causal effects between the two levels of description.

That's a point that we will address with greater extension in a forthcoming paper.

### Acknowledgment

I am indebted to Jean-Paul Delahaye for many enlightening discussions from the beginning of this work, to Serge Grigorieff for having read a first version of this paper and given a counter example that led me to modify the initial definition of an approximation and to build the present one and to Jean-Michel Ghidaglia for a useful discussion on appendix B.

## Appendix A: the asymptotic notations

The asymptotic notations are useful for comparing the order of magnitude of different functions. We recall here the standard notations.

- $f(n) = O(g(n))$ if there are constants $c > 0$, $n_0 > 0$ such that $\forall n > n_0, |f(n)| \leq c|g(n)|$.

- $f(n) = o(g(n))$ if $\lim_{n \to \infty} f(n)/g(n) = 0$.

- $f(n) = \Omega(g(n))$ if there are constants $c > 0$, $n_0 > 0$ such that $\forall n > n_0, |f(n)| \geq c|g(n)|$.

- $f(n) = \omega(g(n))$ if $\lim_{n \to \infty} f(n)/g(n) = \infty$.

- $f(n) \backsim g(n)$ if $\lim_{n \to \infty} f(n)/g(n) = 1$.

- $f(n) = \Theta(g(n))$ if there are constants $c > 0$, $c' > 0$, $n_0 > 0$ such that $\forall n > n_0, \ cg(n) \leq f(n) \leq c'g(n)$

## Appendix B

We prove here that $\sum_{i=1}^{n} F(i))/i = O(F(n))$ if F is a convex function and $F(n) = \Omega(\log n)$.

$$\lim_{n \to \infty} \sum_{i=1}^{n} \frac{F(i)}{i} = \lim_{n \to \infty} \int_{1}^{n} \frac{F(x)}{x} \mathrm{d}x$$

$F(y) = \int_{1}^{y} F'(x)\, dx + F(1)$ pour $y > 1$

Now if F is convex $\int_{1}^{y} F'(x)\, dx \leq (y\text{-}1)\, F'(y) \leq y\, F'(y)$ so

$$\frac{F(y)}{y} \leq F'(y) + \frac{F(1)}{y}$$

$$\int_{1}^{x} \frac{F(y)}{y}\, dy \leq F(x) - F(1) + F(1)\log x$$

Now if $\log x = O(F(x))$ then

$$\int_{1}^{x} \frac{F(y)}{y}\, dy = O(F(x))$$